\documentclass[aps,prb,preprintnumbers,amsmath,amssymb,twocolumn,
               floatfix]{revtex4}
\pdfoutput=1
\usepackage{amsmath}
\usepackage{amssymb}
\usepackage{epsfig}
\usepackage{textcomp}

\begin{document}

\title{Charge-induced conformational changes of dendrimers}

\author{Ronald Blaak}

\author{Swen Lehmann}
\altaffiliation{Current address: Institut f{\"u}r Physikalische Chemie,
RWTH Aachen, Landoltweg~2, D-52056 Aachen, Germany}

\author{Christos N. Likos}

\affiliation{Institut f\"ur Theoretische Physik II: Weiche Materie,
  Heinrich-Heine-Universit\"at D{\"u}sseldorf, Universit\"atsstra{\ss}e 1, D-40225
  D\"usseldorf, Germany} 

\date{\today, submitted to {\sl Macromolecules}}

\begin{abstract}
We study the effect of chargeable monomers on the conformation of dendrimers of low generation by computer simulations, employing bare Coulomb interactions. The presence of the latter leads to an increase in size of the dendrimer due to a combined effect of electrostatic repulsion and the presence of counterions within the dendrimer, and also enhances a shell-like structure for the monomers of different generations. In the resulting structures the bond-length between monomers, especially near the center, will increase to facilitate a more effective usage of space in the outer-regions of the dendrimer.
\end{abstract}

\maketitle 

\section{Introduction}

Dendrimers have been the subject of intensive investigations ever since
their synthesis\cite{vogtle} in the late 1970s. They are characterized 
by a high degree of monodispersity and a
well-defined, highly branched internal structure; efficient
dendrimer
assembly has been boosted by recent progress in synthetic techniques.\cite{antoni:07}
A great deal of research activity has focused on
the issue of whether they possess an open, {\it dense-shell} or 
a collapsed, {\it dense-core} configuration, the motivation arising
by the potential to employ them as hollow, carrier-type molecules
in the former case. For neutral dendrimers, a large number of 
simulation studies,\cite{ballauff:likos:review}
careful self-consistent field calculations\cite{zook} 
and not least scattering experiments\cite{ballauff1,ballauff2} have
revealed that the dense-shell conformation is {\it not} the real one.
Due to back-folding of the end-groups, caused by entropic considerations,
a dense-core calculation results instead, leading even to compact,
hard-sphere-like conformations at high generation numbers.\cite{ingo:mm, rathgeber:jcp}
From the point of view of applications,
this may sound like a disappointing result, as one would like to
have dense-shell molecules. However, seen from the angle of fundamental
research, the growing compactness of dendrimers with increasing 
generation number is very welcome, since it allows to use them
as model colloidal/nano particles with tunable stiffness,\cite{cnl:jcp, ingo:jcp}
bridging the gap between flexible polymers and rigid spheres.

The issue of dendrimer conformations is less clear when {\it charged}
or {\it polyelectrolyte} dendrimers are considered. Charge on the
building blocks of, e.g., poly(amidoamine) (PAMAM)
dendrimers can be manipulated by changing the pH of the solution\cite{nisato}
and their conformations can be further influenced by added salt.
The expectation that charged dendrimers may achieve stretching
is based on experience with other branched polyelectrolytes (PE),
such as PE-stars, for instance.\cite{arben:prl1} However, the
SANS-study of Nisato {\it et al.}\cite{nisato} has led to a
negative result: the size of dendrimers is insensitive to pH
changes and thus to charge. This experimental fact is at odds
with the earlier work by Welch and Muthukumar,\cite{welch}
who predicted, by means of simulation and theory, an `opening up'
of PE dendrimers upon increase of the number of charged units;
similar conclusions were reached in the Brownian Dynamics simulations
of Lyulin {\it et al.}\cite{lyulin} However, both simulational
works quoted above employed a Debye-H{\"u}ckel (screened Coulomb)
interaction potential acting between charged units, treating
thereby the counterions as a continuum. Giupponi {\it et al.},
on the other hand, pointed out that a more realistic treatment
of the molecules should employ the bare Coulomb interactions among
all involved species (monomers and counterions), which is the
method they employed in their own, recent simulations.\cite{giupponi}
Interestingly enough, they found that the dendrimer size is indeed
very weakly dependent on charge, due to local charge neutrality,
a condition that is masked when one employs the Debye-H{\"u}ckel
approximation.

Though the work of Giupponi {\it et al.}\cite{giupponi} has illuminated
a number of issues pertaining to monovalently charged monomer units
accompanied by monovalent counterions, the question of the influence
of valency on dendrimer conformations has not been studied so far.
The purpose of this work is to examine precisely this issue, which
appears relevant on the grounds that counterion valency is known
to bring about drastic changes in, e.g., the size of spherical
PE brushes.\cite{arben:prl2} We focus thereby on dendrimers of
the fourth generation ($G=4$) and examine separately the conformations
of neutral dendrimers, as a reference point, as well as the 
combinations $(Z_m, Z_c) = (1,1), (1,2), (2,1)$ and $(2,2)$,
where $Z_m$ stands for the valency of the monomers and $Z_c$
for the valency of the counterions. We find that the case of
divalent monomers and monovalent counterions brings about a
substantial change of the dendrimer size, accompanied by a strong
stretching of the chemical bonds, whereas the same phenomena are
less pronounced for the other two cases. After describing our
simulation model in Sec.\ II, we present and discuss the results
in Sec.\ III, whereas in Sec.\ IV we draw our conclusions. 

\section{The simulation model}
The dendrimers we use within our simulations are built from a central pair of joined monomers, the so-called generation 0. A successive generation $g+1$ of dendrimer is formed by connecting two additional monomers, the functionality of the dendrimer is therefore three, to each outer monomer of the dendrimer of generation $g$. In doing so, the number of monomers $n(g)$ of a given generation $g$ in a dendrimer follows a simple power law, i.e. $n(g)=2^{g+1}$.

There are in general three types of interactions between monomers. The first type of interaction prevents the collapse of monomers onto each other and is a short-range repulsive interaction given by a simple, shifted and purely repulsive Lenard--Jones potential
\begin{equation}
\label{E:V_LJ}
V_{\text{LJ}} = 
\left \{ \begin{array}{ll}
4 \epsilon \left[ \left( \frac{\sigma}{r} \right)^{12} - \left( \frac{\sigma}{r} \right)^{6} + \frac{1}{4} \right] & r \leq r_c \\
0 & r>r_c 
\end{array} \right.
\end{equation}
where $\sigma$ and $\epsilon$ are the unity of length and energy respectively, and $r_c=2^{1/6}\sigma$ is the range of the interaction.

The second type of interaction is of an attractive nature and describes the bonds between joined monomers in order to prevent the molecule from flying apart. This interaction is described by a FENE 
potential\cite{fene}
\begin{equation}
\label{E:V_FENE}
V_{\text{FENE}} = 
\left \{ \begin{array}{ll}
-15 \epsilon \left( \frac{R_0}{\sigma} \right)^{2} \ln \left[ 1 - \left( \frac{r}{R_0} \right)^{2} \right ] & r \leq R_0 \\
0 & r>R_0
\end{array} \right.
\end{equation}
where $R_0 = 1.5 \sigma$ is the maximum allowed distance between two connected monomers.

The last type of interaction to be included in the model 
is the Coulomb potential between charged monomers
\begin{equation}
\label{E:V_Coul}
V_{\text{Coulomb}} = k_\text{B} T \lambda_\text{B} \frac{Z_i Z_j}{r_{ij}}
\end{equation}
with $Z_i$ and $Z_j$ the charge numbers, $k_\text{B}$ the Boltzmann constant, $T$ the temperature, and $\lambda_\text{B}$ the Bjerrum length given by
\begin{equation}
\label{E:lB}
\lambda_\text{B} = \frac{e^2}{\varepsilon_r k_\text{B} T}
\end{equation}

We have performed molecular dynamics simulations at constant density and temperature using the Nos\'e-Hoover thermostat\cite{Book:Allen-Tildesley,Book:Frenkel-Smit} on dendrimers of generation 4, i.e., a dendrimer formed by 62 identical monomers, where the monomers with charge number $Z_m$ are either neutral, monovalent, or divalent. In order to guarantee charge neutrality the charged dendrimers need to be balanced, for which we in either case used both monovalent and divalent counterions (charge number $Z_c$). Apart from the Coulomb interaction between monomers and counterions, as well as between the counterions themselves, we use for simplicity the same short range repulsion (\ref{E:V_LJ}) as is used for the monomers.

Simulation parameters are chosen such that our unit of length $\sigma = 2.84${\rm \AA}
and we fixed the temperature to $T = 1.2 \epsilon/k_\text{B}$. Using an implicit solvent that mimics the behavior of water, i.e., a temperature of 300K and a relative permittivity $\varepsilon_r = 80$ we arrive at $\lambda_\text{B}/\sigma = 3$. Periodic boundary conditions have been applied in combination with the Ewald summation method to include the long-range electrostatic interactions. The volume of the simulation box was chosen such that effectively the dendrimers can be considered to be independent, indicated by the independence of the results on larger volume sizes. The initial configurations were equilibrated over times long enough for the counterions to diffuse in to the core of the dendrimer, and to reach a steady state for a in- and out-flux. The simulation time is chosen long enough for the individual counter-ions to explore the full dendrimer, i.e., inner and outer ranges.

\section{Results}
In order to characterize the size of the dendrimer, the radius of gyration $R_g$ is measured, which is defined by 
\begin{equation}
R_g^2 = \left \langle \frac{1}{N} \sum_{i=1}^N (\vec{r}_i - \vec{r}_{CM})^2 \right \rangle,
\end{equation}
where the summation runs over the $N$ positions $\vec{r}_i$ of the monomers of the dendrimer and $\vec{r}_{CM}$ is the center of mass of the dendrimer.

The resulting values for the neutral and charged dendrimers are listed in Table ~\ref{T:Rg}. The first observation we can make, is that by charging the dendrimer the radius of gyration increases. This is actually not so surprising, because the monomers will repel each other due to their charges. What is more interesting is the fact that the behavior on charging is not monotonic, i.e., the radius of gyration of the divalently charged monomers and counterions is smaller with respect to that of the monovalent case. This suggests that the swelling of the dendrimer due to the increasing charge is counteracted by the transport of counterions into the deeper regions of the dendrimer, which results in screening effects and reduces the swelling. \cite{Galperin,Gurtovenko,Majtyka}

\begin{table}[thb]
\begin{tabular}{c|l|l}
$Z_m - Z_c$ & $R_g^2/\sigma^2$ & $R_g/\sigma$ \\
\hline
neutral & 6.31 $\pm$ 0.02 & 2.51 $\pm$ 0.01\\
1-1     & 8.20 $\pm$ 0.02 & 2.86 $\pm$ 0.01\\
1-2     & 7.12 $\pm$ 0.02 & 2.67 $\pm$ 0.01\\
2-1     & 9.27 $\pm$ 0.02 & 3.04 $\pm$ 0.01\\
2-2     & 7.49 $\pm$ 0.02 & 2.74 $\pm$ 0.01
\end{tabular}
\caption{The radius of gyration for the neutral dendrimer and the dendrimers with charged monomers and counterions, with charge numbers $Z_m$ and $Z_c$ respectively.}
\label{T:Rg}
\end{table}

It also appears that divalent counterions reduce the size of the dendrimer with respect to that of monovalent counterions. Not only do they result in a better screening of the monomers, but also less of them are required which also leads to less steric hindrance. Using a similar argument, the monovalent monomers reduce the size of the dendrimer with respect of that of divalent monomers, because in the latter case more counterions are required within the dendrimer, preventing  its shrinking. Note that for a neutral dendrimer there is no counterpart for the  counterions present within the dendrimer, hence it is only the short-range repulsion between the monomers that prevents the dendrimer from collapsing.

\begin{figure}[thb]
\centering
\includegraphics[width=8cm]{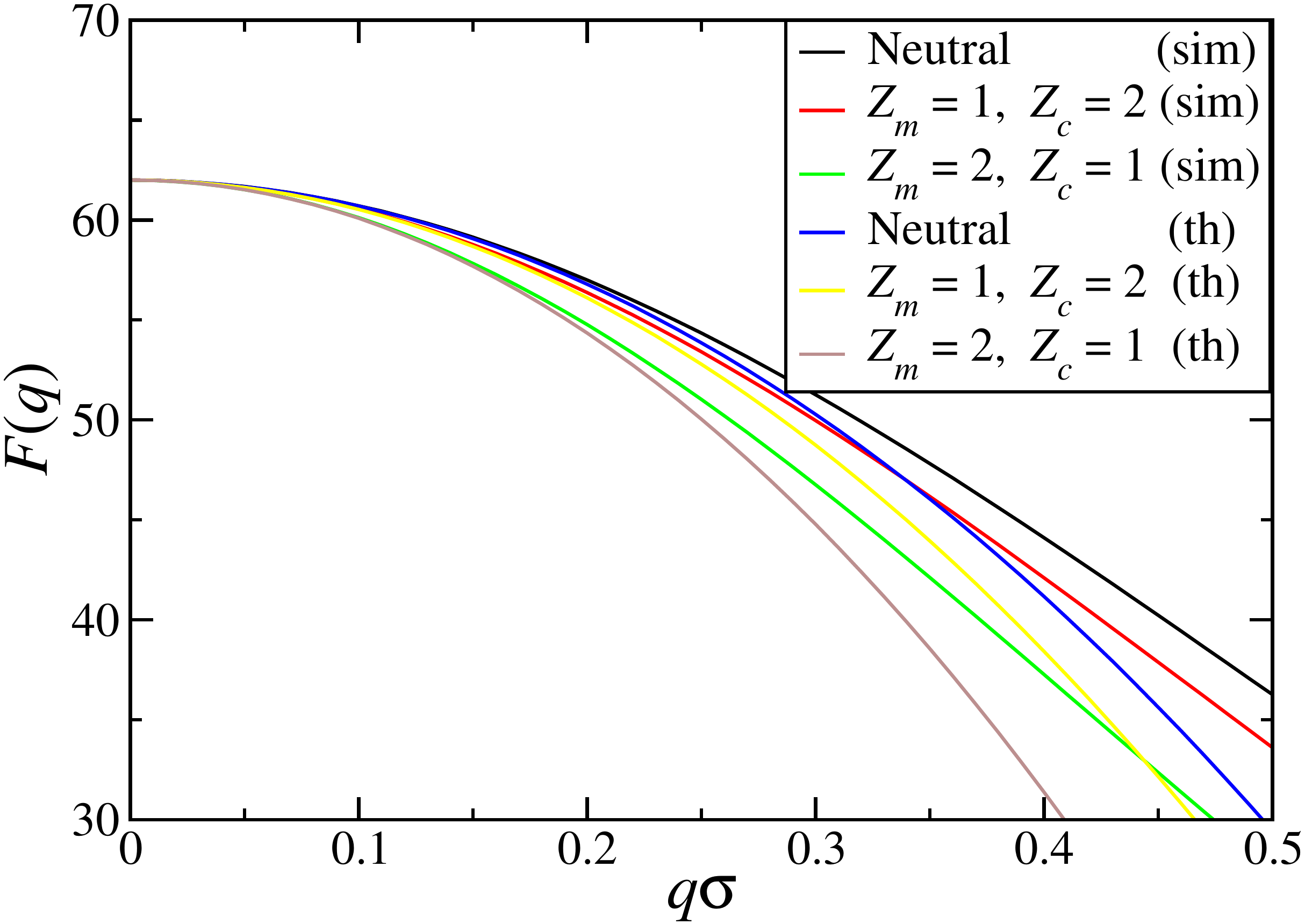}
\caption{Comparison of the measured form factor, Eq.~(\ref{E:Fq})
from the simulation (sim), with the small wave-vector limit, Eq.~(\ref{E:Fq-small}), 
denoted (th).}
\label{F:Fq}
\end{figure}

The radius of gyration is also experimentally observable from the small wave-vector limit of the form factor $F(\vec{q})$ defined by
\begin{equation}
\label{E:Fq}
F(\vec{q}) = 1 + \frac{1}{N} \left \langle \sum_{i \neq j} \exp(-\imath \vec{q} \cdot \vec{r}_{ij})\right \rangle.
\end{equation}
By expanding in small wave-vectors $q$ and averaging over the directions it is easy to show that this results in
\begin{equation}
\label{E:Fq-small}
F(q) = N \left[1 - \frac{(q R_g)^2}{3} \right].
\end{equation}
In the present case this expansion is valid up to $q \sigma \approx 0.2$ as can be seen in Fig.~\ref{F:Fq} where we compare the measured form factor,
Eq.~(\ref{E:Fq}), from the simulation with the small wave-vector limit,
Eq.~(\ref{E:Fq-small}).

\begin{figure}[thb]
\centering
\includegraphics[width=8cm]{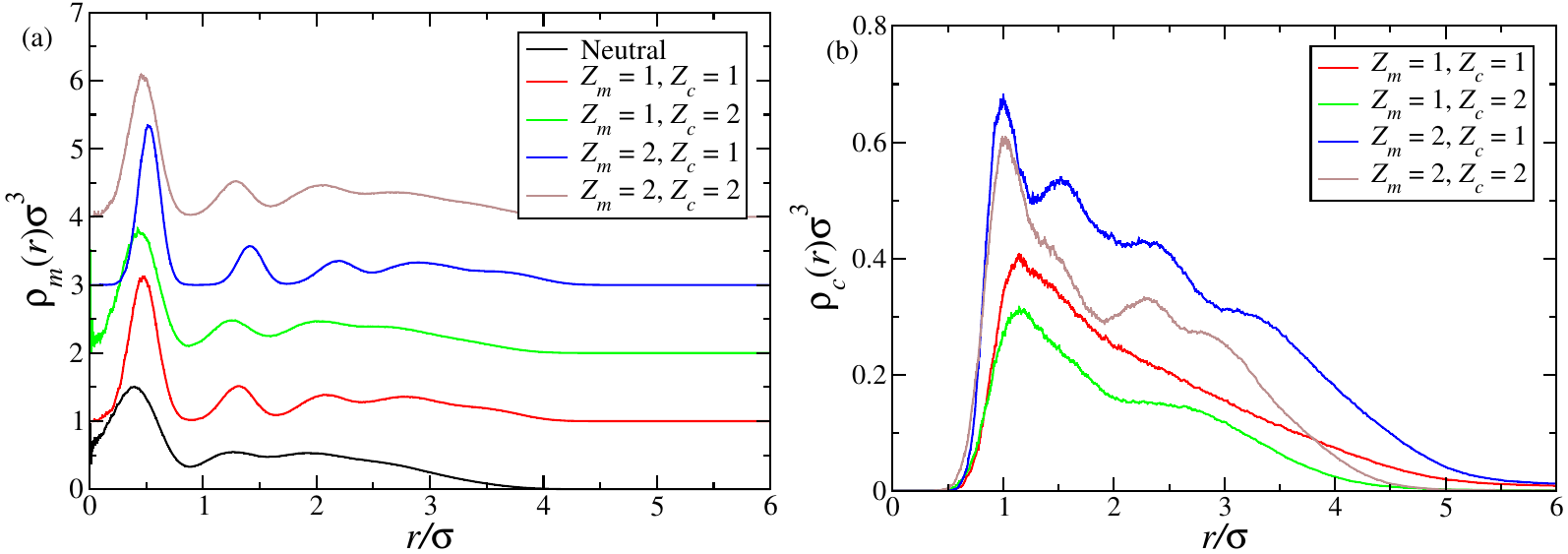}
\caption{Radial density profiles for the monomers (a) and counterions (b), measured with respect to the center-of-mass of the dendrimer. The curves in the left-hand figure have been shifted to facilitate a comparison.}
\label{F:Rho}
\end{figure}

In Fig.~\ref{F:Rho} the radial density profiles of the monomers and counterions are shown, both measured with respect to the center of mass of the dendrimer. Note that at small distances from the center of mass the noisy behavior is purely due a poor statistical sampling caused by the lack of particles in that region.

It is immediately clear that the presence of charge on the monomers results in a much more structured density profile, in which the monomers are mostly found in a shell-like structure. The mutual repulsion of the monomers due to their charge not only leads to a larger size of the dendrimer but it enables the counterions to diffuse into the dendrimer as well; this can be seen from their density profiles. Their distribution, however, shows less structure and has a wider range. The latter can also easily be understood, since the presence of counterions within the dendrimer will also lead to steric hindrance and there is an obvious entropic gain in surrounding the dendrimer and move in the region lying
outside its extent.

\begin{figure}[thb]
\centering
\includegraphics[width=8cm]{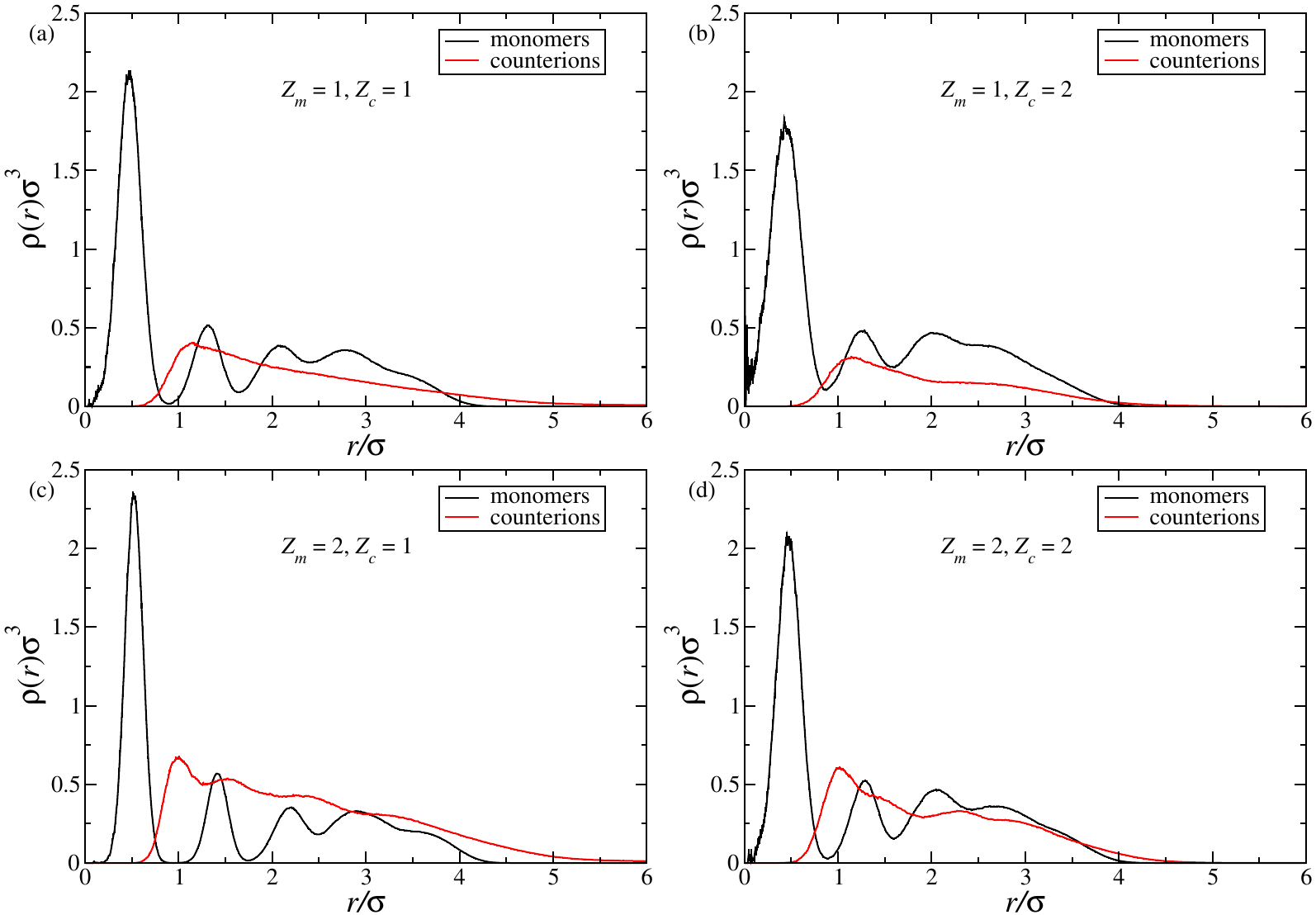}
\caption{Radial density profiles of the monomers and counterions for the monovalently [(a),(b)] 
and divalently [(c),(d)] charged monomers, 
and monovalently [(a),(c)] and divalently [(b),(d)] counterions.}
\label{F:Rho_m_c}
\end{figure}

To clarify the ordering in the charged dendrimer cases, the same data are shown in Fig.~\ref{F:Rho_m_c} where the radial density profiles of the monomers and counterions are directly compared. Although the highest density for the counterions is reached roughly in between the first and second peak of the monomer distributions (except for the case $Z_m=1$ and $Z_c=2$), there is no layered structure present for the counterions. This is also confirmed by the absence of plateaus in the cumulative counterion density profile (not shown).

\begin{figure}[thb]
\centering
\includegraphics[width=8cm]{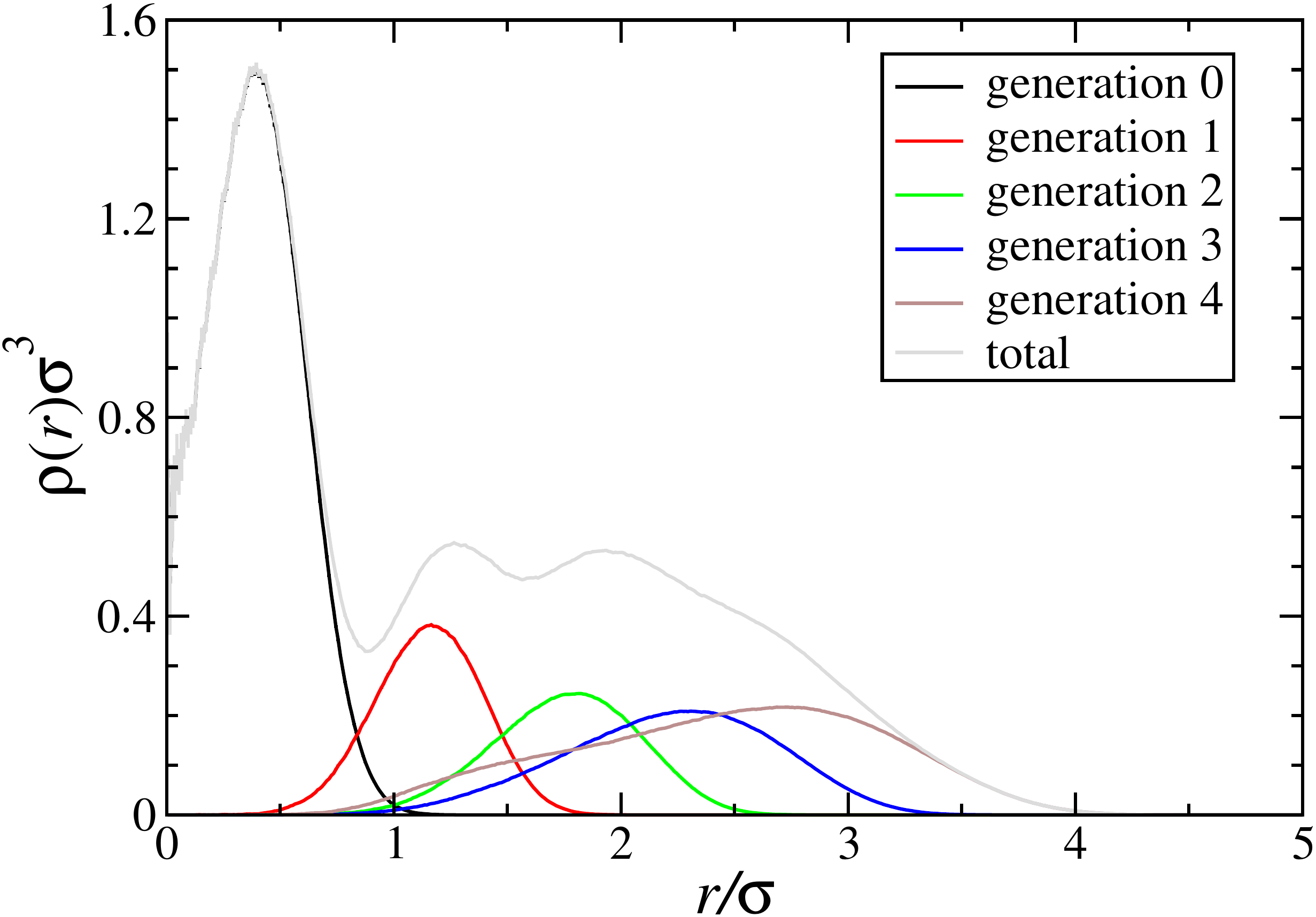}
\caption{Radial density profiles of the monomers in the neutral dendrimer decomposed in the contributions stemming from each generation.}
\label{F:Rhog_00}
\end{figure}

The absence of charges on the monomers in the neutral dendrimer does not only affect its size, i.e., it is more compact, but it also modifies its internal structure as is illustrated in Fig.~\ref{F:Rhog_00}, where the density profiles for the different generations are plotted versus the distance to the center of mass. This reveals that the density of generation 4 monomers at smaller distances is larger than that of those from generation 3. In other words, the dendrimer starts to fold in to itself. Whereas the repulsive Coulomb interactions between charged monomers tend to stretch the dendrimer causing a loss in entropy, the neutral monomers just feel the short-range Lennard-Jones repulsion and can exploit the open space in the core of the dendrimeric structure. A not unimportant other reason is the absence of the equivalent of counterions within the dendrimer. This effect is only weakly visible in the cases of the divalently charged counterions, but it is to be expected that it will be more pronounced for increasing value of the generation of the dendrimer.

\begin{figure}[thb]
\centering
\includegraphics[width=8cm]{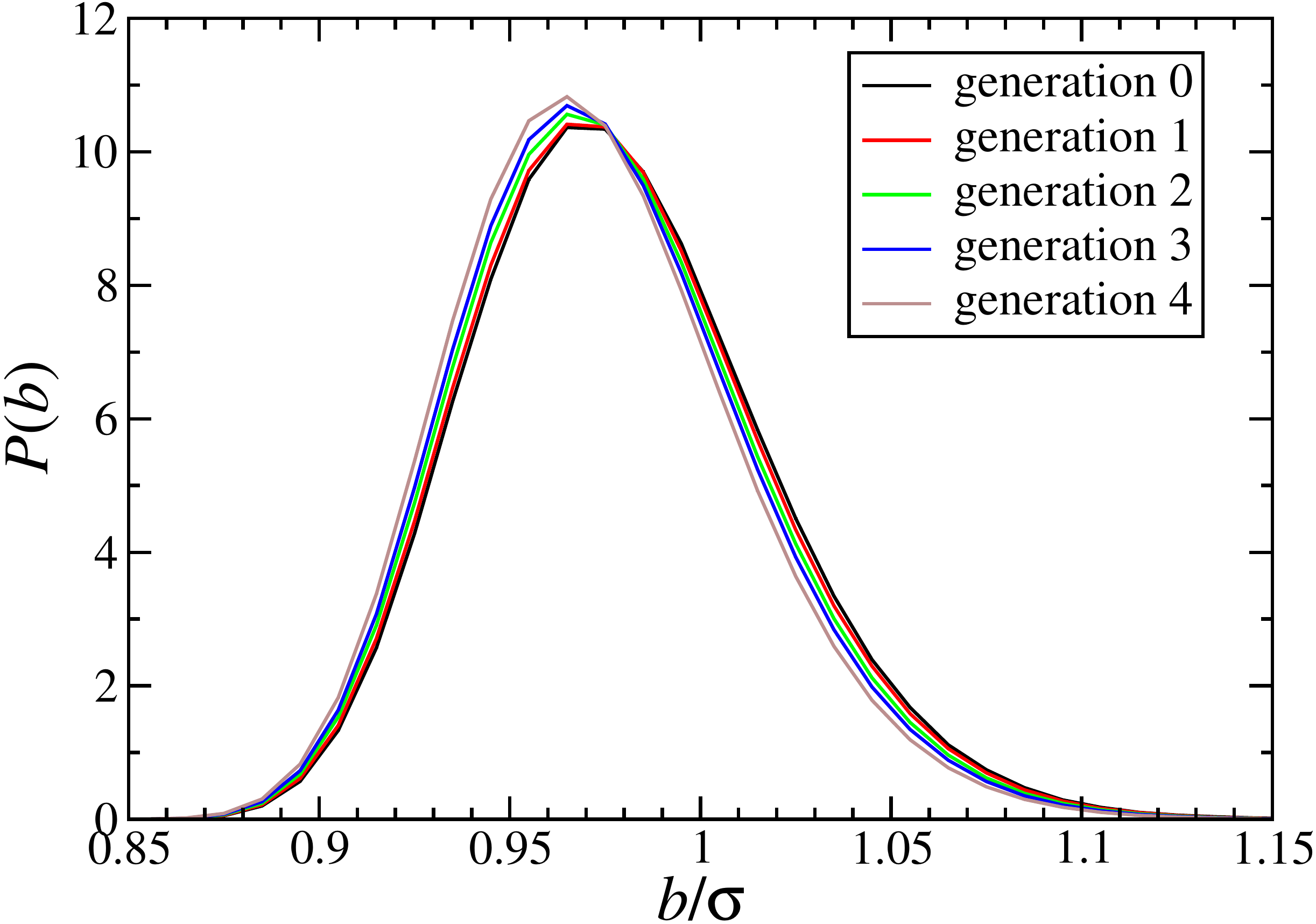}
\caption{Probability distribution of bond-lengths for each generation for the neutral dendrimer.}
\label{F:Bonds_00}
\end{figure}

\begin{figure}[thb]
\centering
\includegraphics[width=8cm]{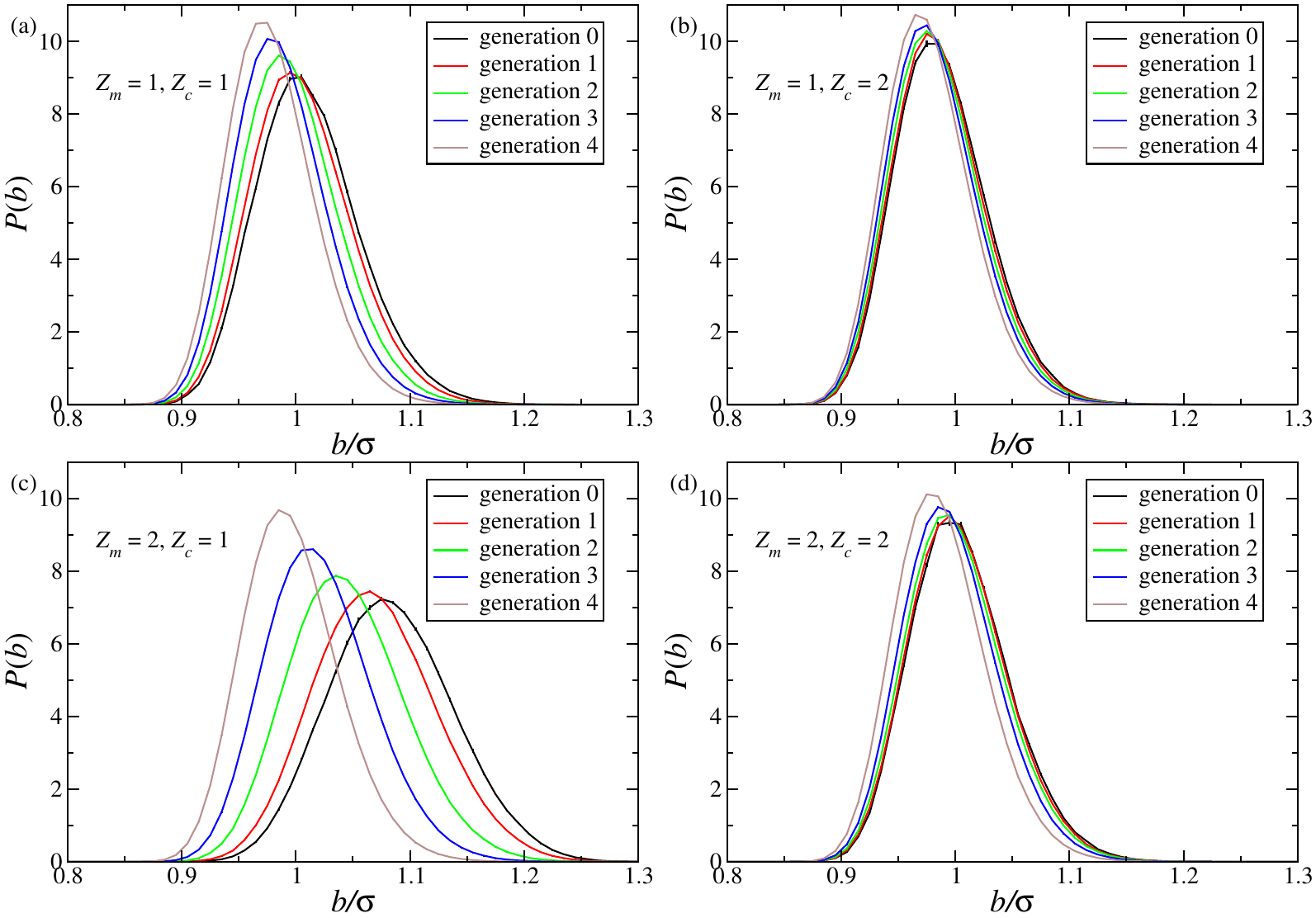}
\caption{Probability distribution of bond-lengths for each generation for the for the 
monovalently [(a),(b)] and divalently [(c),(d)] charged monomers, 
and monovalently [(a),(c)] and divalently [(b),(d)] charged counterions.}
\label{F:Bonds_XX}
\end{figure}

A more detailed description of the internal structure of the dendrimer can be obtained by analyzing the the bond-lengths $b$, shown in Figs.~\ref{F:Bonds_00} and \ref{F:Bonds_XX}. The bond-length probability distributions, $P(b)$, are decomposed per generation, whereby a bond of generation $n$ is formed by a monomer of generation $g$ with its parent of generation $g-1$ (with the exception of generation 0). For the neutral dendrimer only a small shift in the distribution of lengths is found towards shorter bond-lengths for higher generations, i.e., bonds near the center of the dendrimer tend to be more stretched than those near the border. This implies that there is a collective behavior in which the mutual repulsion of monomers in higher generations that prefers to expand the dendrimer, forcing the fewer central bonds to stretch.

\begin{figure}[thb]
\centering
\includegraphics[width=8cm]{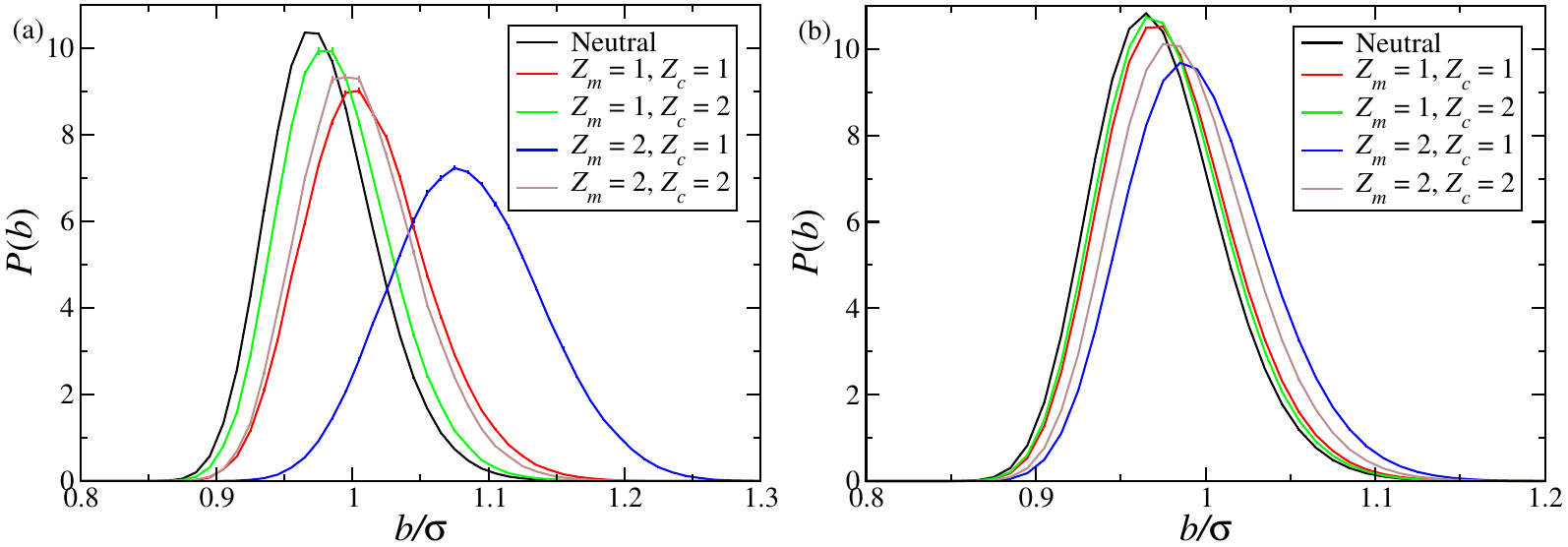}
\caption{Probability distribution of bond-lengths of generation 0 (a) and generation 4 (b) 
for the different cases monomers and counterions.}
\label{F:Bonds_x}
\end{figure}

Fig.~\ref{F:Bonds_XX} shows the same distributions but now for the charged dendrimer cases. The first thing one can observe is that the bond-lengths in the charged cases are more stretched. This is not so surprising, since the bond-length is directly affected by the mutual charge repulsion of the monomers and even more so for the divalent monomers. Also the tendency for the central bonds to be more stretched than those in the outer regions is apparent. The most interesting case, however, is that of the divalent monomers and monovalent counterions, in which case the stretching of the bond-lengths shifts significantly with the generation. In Fig.~\ref{F:Bonds_x}  this is illustrated even more clearly by comparing the bonds of generation 0 and 4 for the various models. This suggests that the abundance of monovalent counterions required for an effective screening of the divalent monomer bonds can not be packed within the core of the dendrimer.

\begin{figure}[thb]
\centering
\includegraphics[width=4cm]{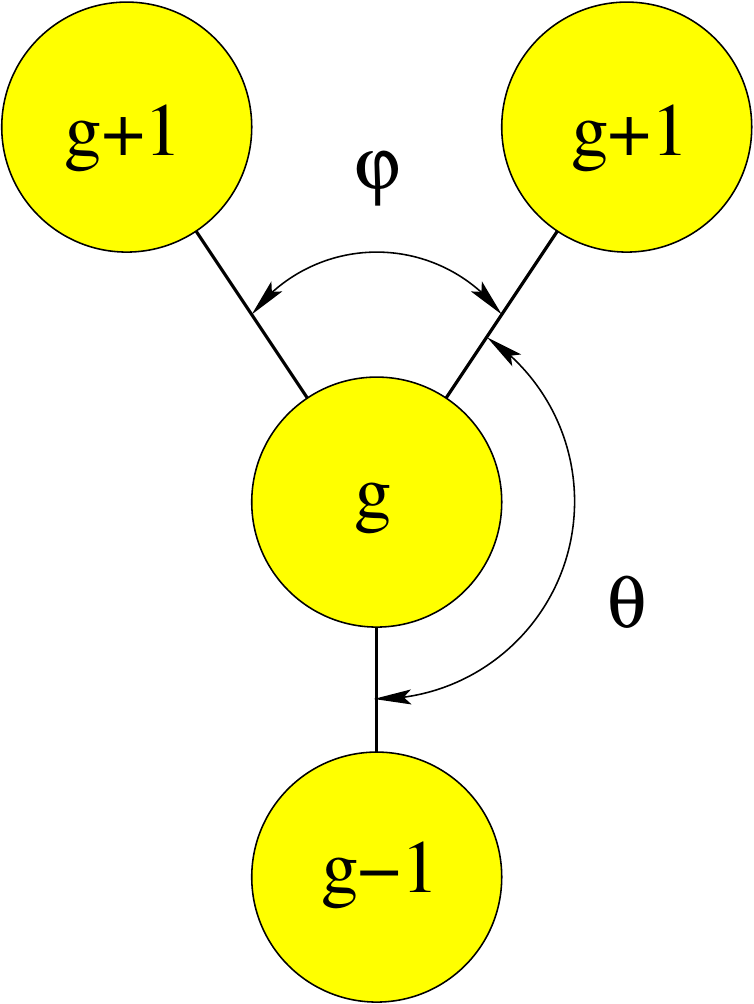}
\caption{Schematic representation of the local structure of a monomer inside a dendrimer indicating the different $\theta$ and $\phi$ angles mentioned in the text.}
\label{F:Angle}
\end{figure}

In the final method we used to examine the internal structure of the dendrimer, we consider the angles between bonds in the dendrimer. The angle $\phi$ as illustrated in Fig.~\ref{F:Angle} is the angle between the bonds of a monomer of generation $g$ and the two monomers bounded to it of generation $g+1$ (a bond generation $g+1$). The angle $\theta$ is the angle between a bond of generation $g$ and a bond of generation $g+1$.

\begin{figure}[thb]
\centering
\includegraphics[width=8cm]{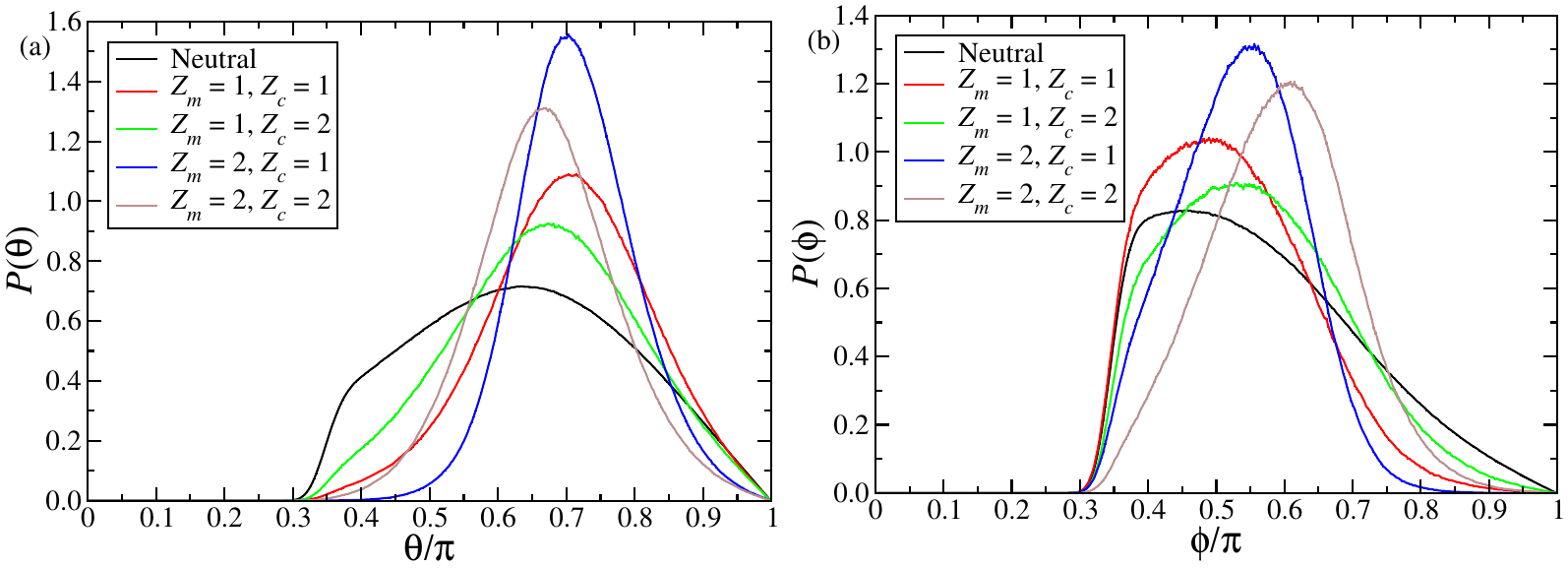}
\caption{Probability distribution of the angles $\theta$ [(a)]
and $\phi$ [(b)] for the different dendrimer cases averaged over all nodes in the dendrimer.}
\label{F:Angles_m}
\end{figure}

The results of the angle analysis are shown in Fig.~\ref{F:Angles_m}, where we have averaged over all 30 nodes in the dendrimer. In the case of the neutral dendrimer the probability distributions are broader than in the case of the charged dendrimers. This is partially due to the fact that there is no penalty for smaller angles, whereas in the charged dendrimer cases the angles will be more focused in the equal angles area due to the Coulomb repulsion.

\begin{figure}[thb]
\centering
\includegraphics[width=8cm]{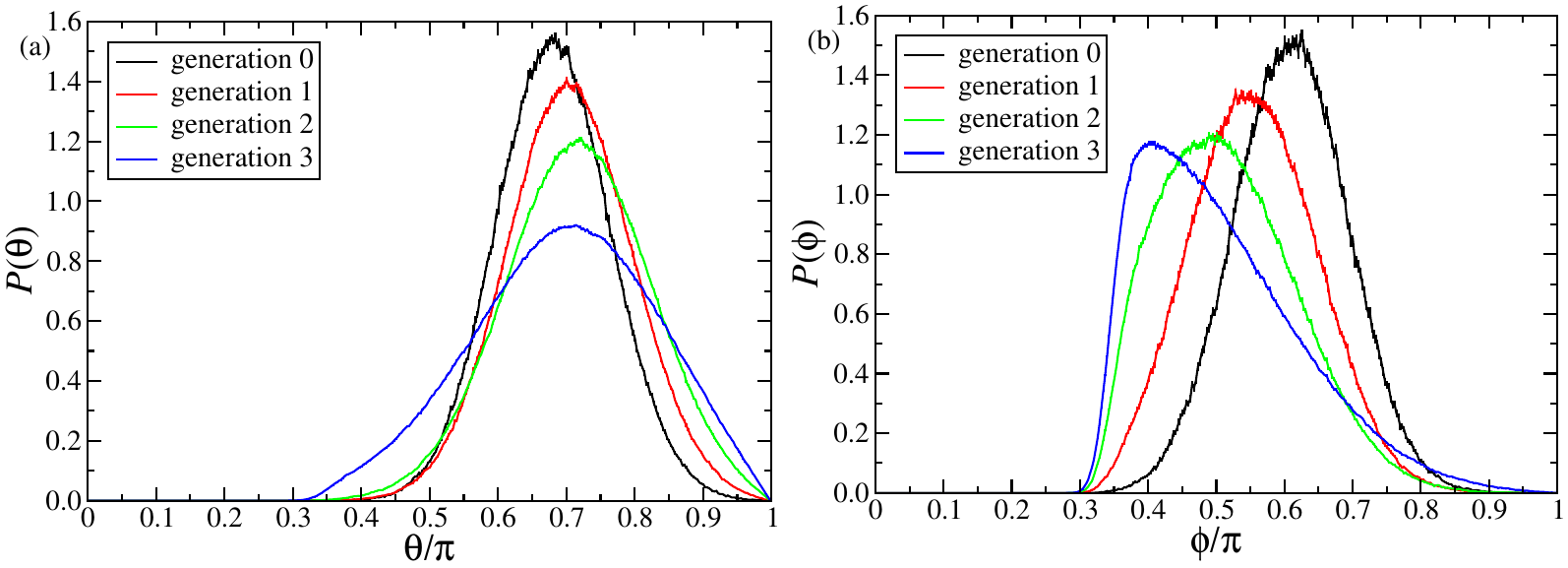}
\caption{Probability of the angle distributions of a dendrimer with 
monovalent monomers and counterions. (a): angle $\theta$; (b): angle $\phi$.}
\label{F:Angles_MM}
\end{figure}

Also these angle distributions will, just like in the case of the bond-lengths, depend on the position of the monomer within the dendrimer, i.e. its generation, as can be seen in Fig.~\ref{F:Angles_MM} in the case of monovalent monomers and dendrimers. For nodes near the core the $\phi$ angles are more or less centered about 120\textdegree, but for nodes near the boundary of the dendrimer the peaks shifts to smaller angles. This is specially clear for the angle $\phi$ for the outer most nodes lies between 70\textdegree~ and 80\textdegree~ for the various cases presented here. At the same time the  $\theta$ angles show a similar behavior, except that they become larger, albeit that the shift is not as significant as is the case for $\phi$.

This is not the only information that can be extracted from the angles $\theta$ and $\phi$. In general the three bonds at a given node will not lie within a single plane, but span a space angle, also called spherical excess,\cite{Book:Korn-Korn} $\varepsilon$, 
which ranges from zero to $2\pi$. Both extremes can only be obtained in the limiting case of three bonds lying within the same plane. The fact that for such a planar configuration the spherical excess can take two values can be understood by realizing that two cases need to be distinguished. If the sum of the three angles between the pairs of bonds $\theta_1+\theta_2+\phi=2\pi$, the space angle reaches it maximum $2\pi$ and spans a half sphere. In the other case the largest of the three bond angles equals the sum of the smaller ones, e.g. $\theta_1+\theta_2=\phi$ and the spherical excess reaches it minimum 0. In other words, when the three bonds completely open up and stretch away from
one another, one has the maximum value, $\varepsilon = 2\pi$; 
on the other hand, when there is complete backfolding of the two bond 
connecting the $g$-monomer with the two $g+1$-monomers towards the bond 
between the $g$-monomer and the $g-1$-monomer in Fig.\ 8, 
then $\varepsilon = 0$. In this fashion, the probability distribution 
$P(\varepsilon)$ of the spherical excess offers valuable information on the
presence of back-folding within the dendrimer. A distribution with strongly
suppressed values around $\varepsilon = 0$ and high values around
$\varepsilon = 2\pi$ implies a stretched dendrimer with planar three-bond
junctions at the branching points. Relatively flat distributions with 
non-negligible values around $\varepsilon = 0$ point rather to strongly
back-folded dendrimers, akin to neutral ones.

\begin{figure}[thb]
\centering
\includegraphics[width=8cm]{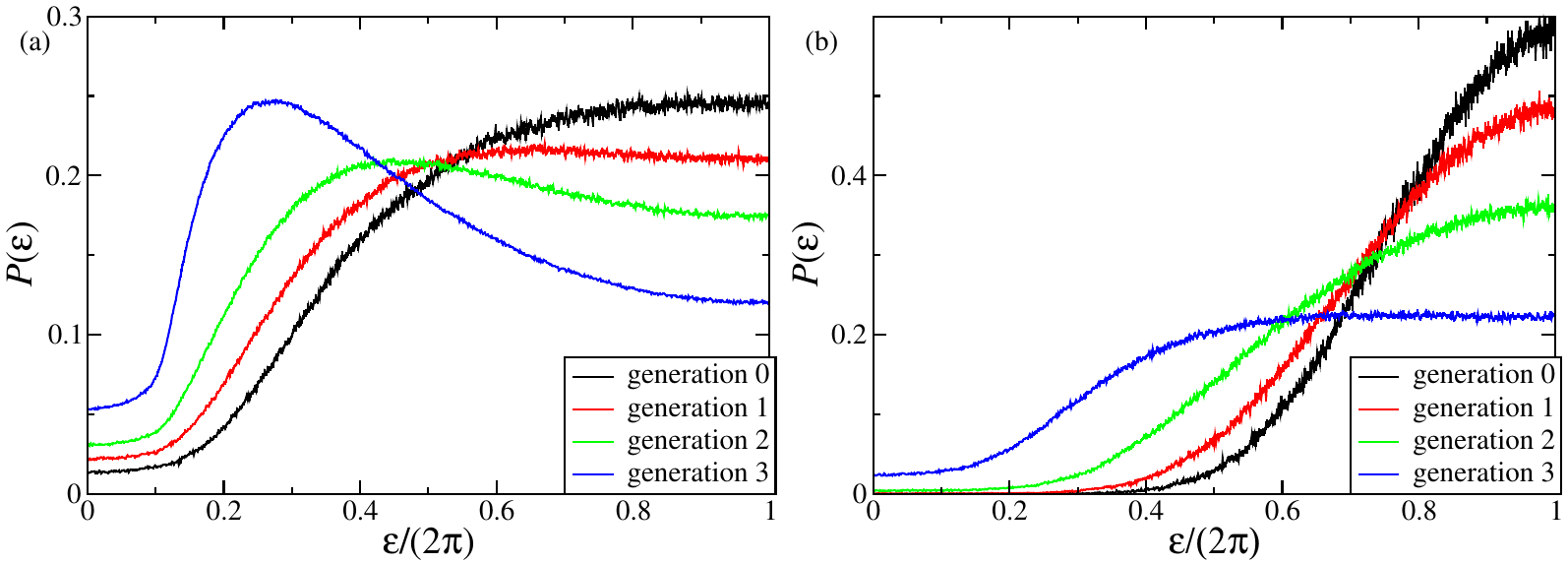}
\caption{Spherical excess for the neutral dendrimer [(a)] and 
monovalently charged monomers and counterions [(b)].}
\label{F:Sph_exc}
\end{figure}

The probability distribution of the spherical excess for the neutral dendrimer and for the case of a charged dendrimer with both monomers and counterions monovalently charged, is shown in Fig.~\ref{F:Sph_exc}. It reveals that the spherical excess decreases for nodes that lie further away from the core of the dendrimer. The preference for spherical excess to be large near the center of the dendrimer can be understood by realizing that such  behavior will maximize stretching in the dendrimer and enable to exploit the so available space for the more distant monomers more effectively. The decrease of the spherical excess is particular striking in the case of the neutral dendrimer and is a necessary consequence from our earlier observation that the monomers from the last generation can fold back into the dendrimer.\cite{ingo:mm}
As can be seen in Fig.~\ref{F:Sph_exc}(b), however, and in conjunction with
Fig.~\ref{F:Sph_exc}(a), the $Z_m = Z_c = 1$-dendrimer shows a considerably
higher degree of stretching than the neutral one; its distribution of the
spherical excess is shifted towards higher values and only at the third
generation do we observe a nonvanishing contribution around $\varepsilon = 0$.

\begin{figure}[thb]
\centering
\includegraphics[width=8cm]{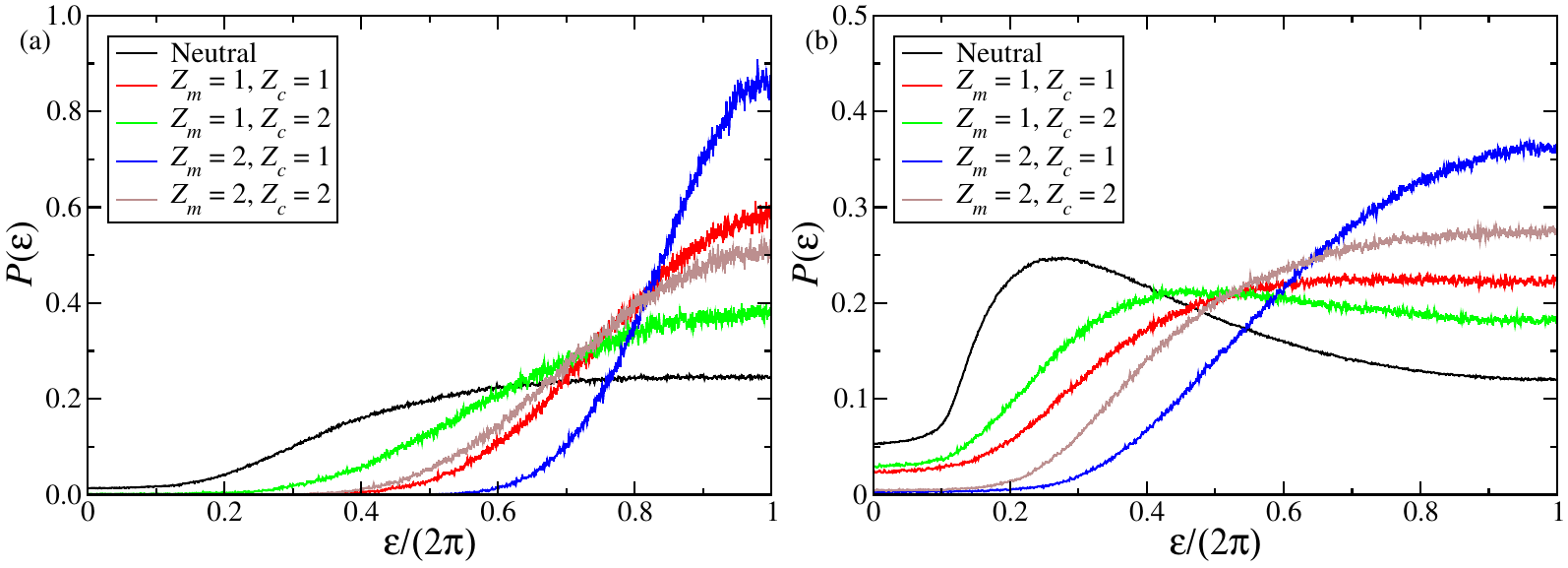}
\caption{Spherical excess of generation 0 [(a)] and generation 3 
[(b)] nodes for the different cases monomers and counterions.}
\label{F:Sph_exc_g}
\end{figure}

Fig.~\ref{F:Sph_exc_g} shows the spherical excess of the nodes 
of generation 0 and 3 for the various models discussed. In full agreement
with all our previous measures, we find that charged dendrimers carrying
divalent monomers and monovalent counterions show the most dramatic
degree of stretching, all the way from the inner to the outermost 
generations. Physically, this is caused by the strong attraction between
the monomers and the counterions. In this case, there are twice as
many counterions as in the cases $Z_m=Z_c=1$ and $Z_m=Z_c=2$
that have to be accommodated in the inner region of the dendrimer, giving rise
to stretching to create space. Two counterions per monomer are 
required to achieve local charge neutrality,\cite{giupponi} rendering
the composite, monomer-counterion groups too big to allow for the
backfolding seen for neutral dendrimers.

\section{Discussion}
We have presented simulation data of simple,
fourth generation dendrimers and investigated the effect that 
chargeable monomers will have on their size and internal structure.
In particular, we have allowed for the possibility of divalent
monomer units and/or counterions and monitored the conformational
changes induced by the various scenarios. For the simplest case,
$Z_m = Z_c = 1$, our findings are in full agreement with the
recent ones of Guipponi {\it et al.},\cite{giupponi} in the sense
that no dramatic changes in the dendrimer conformation and size
have been detected. This is also in agreement with the experimental
results of Nisato {\it et al.}\cite{nisato} and at odds with the
prediction of Welch and Muthukumar.\cite{welch} At this point,
it appears that the assumption of linear counterion screening,
which has been employed in Ref.~\onlinecite{welch} in the form of
a Debye-H{\"u}ckel approximation, is not valid when strong 
counterion absorption and condensation effects are present.
More so at the nanoscale-distances involved in the inner of a
dendrimer.

At the same time, we could establish that a remarkable change of
the size of the dendrimer, accompanied by a stiffening and 
stretching of its bonds, takes place when the former carries
divalent chargeable groups, which release two monovalent counterions
per site. Here, although the resulting configuration is not
yet of the dense-shell type, the dendrimer size grows by almost
50\% in comparison to a neutral dendrimer and strong correlations
between the monomer- and counterion profiles appear. It is now
reasonable to assume that, given a guest molecule that carries
the same type of charge as the counterions and being small enough
to `fit' inside the stretched dendrimer, absorption will take
place, as in this fashion a number of counterions will be 
released from the inner part of the dendrimer, resulting in 
a concomitant entropy 
gain.\cite{grun:04,radler:97,verma:97,evans:03,wittemann:pccp:03,kon}
Another possibility is to endow the dendrimers with a spacer
length $s > 1$, which allows for creation for more space in
their interior but at the same time it also increases the
possibilities to backfolding.
Work along these lines is currently in progress.

\begin{acknowledgments}
This work has been supported by the Deutsche Forschungsgemeinschaft (DFG).
\end{acknowledgments}

\end{document}